\documentclass[12pt]{iopart}

\def\etcl{$\kappa$-(BE\-DT\--TTF)$_2$\-Cu\-[N\-(CN)$_{2}$]Cl}

\def\CuCN{$\kappa$-(BE\-DT\--TTF)$_2$\-Cu$_2$\-(CN)$_{3}$}
\def\AgCN{$\kappa$-(BE\-DT\--TTF)$_2$\-Ag$_2$\-(CN)$_{3}$}
\def\EtMe{$\beta^{\prime}$-EtMe$_3$\-Sb\-[Pd(dmit)$_2$]$_2$}
\def\cm{cm$^{-1}$}

\usepackage[dvips]{graphicx}
\usepackage{graphicx}
\usepackage{color}

\begin{document}

\title[Electrodynamics of quantum spin liquids]
{Electrodynamics of quantum spin liquids}

\author{Martin Dressel and Andrej Pustogow}

\address{1. Physikalisches Institut, Universit\"at Stuttgart, Pfaffenwaldring 57, 70550 Stuttgart, Germany}

\date{\today}

\begin{abstract}
Quantum spin liquids attract great interest due to their exceptional magnetic properties characterized by the absence of long-range order down to low temperatures despite the strong magnetic interaction.
Commonly, these compounds are strongly correlated electron systems, and their electrodynamic response is governed by the Mott gap in the excitation spectrum.
Here we summarize and discuss the optical properties
of several two-dimensional quantum spin liquid candidates.
First we consider the inorganic material Herbertsmithite ZnCu$_3$(OH)$_6$Cl$_2$ and related compounds, which crystallize in a kagome lattice. Then we turn to the organic compounds
$\beta^{\prime}$-EtMe$_3$\-Sb\-[Pd(dmit)$_2$]$_2$,
$\kappa$-(BEDT-TTF)$_2$Ag$_2$(CN)$_3$ and $\kappa$-(BEDT-TTF)$_2$Cu$_2$(CN)$_3$, where
the spins are arranged in an almost perfect triangular lattice, leading to strong frustration.
Due to differences in bandwidth, the effective correlation strength varies over a wide range, leading to a rather distinct behavior as far as the electrodynamic properties are concerned. We discuss the spinon contributions to the optical conductivity in comparison to metallic quantum fluctuations in the vicinity of the Mott transition.
\end{abstract}


\maketitle

\section{Introduction}
\label{sec:introduction}
The pioneering work on geometrical frustration dates back to the 1920s,
when Pauling \cite{Pauling35} realized that the hydrogen bonds between H$_2$O molecules
in ice can be allocated in multiple ways. Any given oxygen atom in
water ice is situated at the vertex of a diamond lattice and has four
nearest-neighbor oxygen atoms, each connected via an intermediate
proton. According to the ice rule, the
lowest energy state has two protons positioned close to the oxygen
and two protons positioned farther away, forming a ``two-in two-out''
state. Although these considerations dealt with electric dipoles, Anderson
\cite{Anderson56,Anderson73,Fazekas74} mapped them to a spin model possessing an extensive degeneracy of states. Two-dimensional arrangements and many generalizations
have been studied subsequently \cite{Lee08,Balents10,FrustratedMagnetism11,FrustratedSpinSystems13,Starykh15}.

These so-called resonating-valence-bond (RVB) states do not feature
any long-range magnetic order or broken lattice symmetries,
but are believed to exhibit non-local, topological order
\cite{Wen02}.
They can be considered as variational ground states
constituted of only the shortest possible valence bonds (singlets), with equal weights
for all bond configurations.
While the formation of a valence bond implies a gap to excite those two spins,
long-range valence bonds are more weakly bound and thus a gapless spectrum is possible in spin-$\frac{1}{2}$ systems.
It was suggested that such states can be described by gauge theories and
these gauge excitations should be visible in the spectrum \cite{Savary17}.
In his seminal paper of 1987 Anderson applied the idea of resonating valence bonds to high-temperature superconductivity in cuprates \cite{Anderson87,Baskaran87,Zaanen12}.

Pure dipolar interaction has been realized in fermionic and bosonic quantum gases, where
large magnetic moments provide the dominant interaction \cite{Lahaye09,Baranov12}
in the absence of any other force.
In solids, however, exchange and Ruderman-Kittel-Kasuya-Yosida (RKKY) interactions are of superior importance \cite{Belobrov83,Johnston16}, and there are only few examples of mainly dipolar lattices, such as isolated water molecules in nano-pores \cite{Gorshunov13,Gorshunov16}.

\begin{figure}[h]
\centering
\includegraphics[width=0.7\columnwidth]{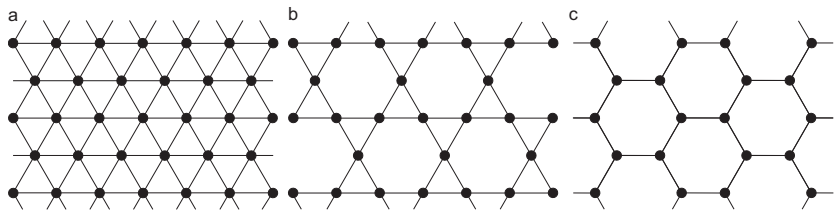}%
\caption{Two-dimensional lattice structures with a high degree of frustration. (a) Triangular lattice, (b) kagome lattice, and (c) hexagonal lattice, where the number of neighbors is reduced from $z=6$, to 4 and down to 3.}%
\label{fig:lattice}%
\end{figure}
While in chains, rectangular and cubic lattices the alternating formation of dipoles can be easily obtained, geometrical frustration becomes an issue in triangular, hexagonal, kagome and
hyper-kagome lattices or tetragonal structures, for instance (figure~\ref{fig:lattice}).
The situation is more intriguing in the case of
quantum spin systems when geometrical frustration and quantum
fluctuations may prohibit the formation of long-range order even at
the lowest temperatures. In these cases liquid-like ground states are
expected, and the numerous investigations of quantum spin liquids and
spin ice reflect these efforts \cite{Nisoli13,Gingras14,Savary17,Zhou17}.

It took decades before the theoretical concept of quantum spin liquids was actually realized in solids,
first in the organic compound \CuCN, which crystallizes in a triangular pattern \cite{Shimizu03,Kurosaki05},
and later in the kagome lattice of ZnCu$_3$\-(OH)$_6$Cl$_2$,
\cite{Shores05,Mendels07,Helton07}. Recently, three-dimensional pyrochlore materials,
such as $A_2B_2$O$_7$ with $A={\rm Pr}$ or Ho and $B={\rm Hf}$, Ti, Zr, etc.,
have been discussed as possible quantum spin liquids with frustration and disorder
of superior importance \cite{Nisoli13,Gingras14,Savary17,Savary17b}.
Still there is a lack of direct experimental confirmation, and also the theoretical description of real systems is unsatisfactory; by now the nature of the spin liquid state must be considered as rather unclear. Is the presence of geometrical frustration sufficient
to realize a spin liquid? What is the influence of disorder, always present in actual materials?
These compounds constitute layered structures, but how important is the coupling to the third dimension?
Naturally, quantum spin liquids are mainly considered from the magnetic point of view and the present status is summarized in several recent reviews \cite{Savary17,Zhou17,Norman16,Kanoda11,Powell11}. But how does the electronic degree of freedom is affected by the formation of a quantum spin liquid? Is it caused by the coupling of spin and charge degrees of freedom, or is it due to geometrical frustation and inherent disorder?
Here we want to focus on the electrodynamic properties of two-dimensional quantum spin liquids and review the experimental facts observed by now.

\section{Herbertsmithite}
\begin{figure}[h]
\centering
\includegraphics[width=0.4\columnwidth]{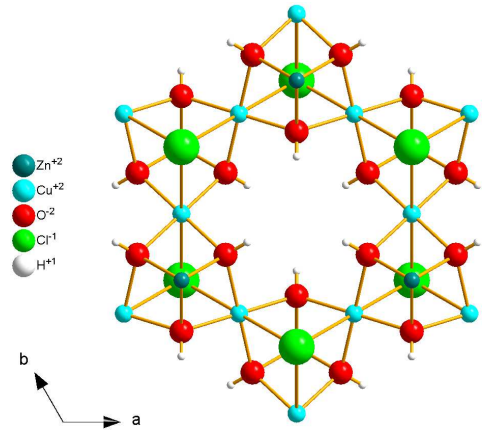}
\caption{The crystal structure of ZnCu$_3$(OH)$_6$Cl$_2$
exhibits the characteristic kagome arrangement of the Cu atoms (cyan) linked by superexchange.}
\label{fig:HS-structure}%
\end{figure}

Among the inorganic spin-liquid candidates, the rhombohedral Herbertsmithite ZnCu$_3$(OH)$_6$Cl$_2$
scored highest in popularity. As  depicted in figure~\ref{fig:HS-structure}, the copper ions form an almost perfect $S=\frac{1}{2}$ kagome lattice \cite{Shores05}, {\it i.e.} corner sharing triangles in the plane
with strong antiferromagnetic superexchange  $J$ of approximately 200~K.
No magnetic order is detected all the way down to $T=50$~mK \cite{Mendels07,Helton07}.
Among the transition metal oxides also other candidates such as the V{\'e}signi{\'e}ite  BaCu$_3$V$_2$O$_8$(OH)$_2$ \cite{Okamoto09} and Volborthite Cu$_3$V$_2$O$_7$(OH)$_2\cdot 2$H$_2$O \cite{Hiroi01,Fukaya03}, and also the hyper-kagome Na$_4$Ir$_3$O$_8$ \cite{Okamoto07} or  PbCuTe$_2$O$_6$ \cite{Khunitia16} have come under scrutiny.
Still the Zn substituted Cu$_4$(OH)$_6$Cl$_2$ remains the prime candidate for
a quantum spin liquid.
Specific heat and inelastic neutron scattering experiments did not find
indications of a spin gap separating $S=0$ and $S=1$ down to 0.1~meV, inferring that the spin excitations form a continuum \cite{Helton07,deVries08,Nilsen13,Han12,Han16}. This important issue, however, is far from being settled, neither from the experimental nor from the theoretical side \cite{Lecheminant97,Sindzingre09,Yan11,Lauchli11,Nakano11,Potter13}.
Among other suggestions \cite{Kalmeyer87,Wen02,Sheng09,Qi09,Mishmash13}, it was proposed that a gapless U(1) spin-liquid state forms with a spinon Fermi surface and with a Dirac-fermion excitation spectrum of the kagome lattice \cite{Motrunich05,Ran07}.

As far as the charge degrees of freedom are concerned, Herbertsmithite exhibits anisotropic electronic properties due to its layered structure. The strong Coulomb repulsion $U\approx 7$ to 8~eV \cite{Pustogow17} makes them clear-cut insulators with vanishing electronic conduction in all directions up to room temperature. Optical transmission and reflection measurements reveal a charge-transfer band around 3.3~eV, corresponding to 26\,600~\cm; between 1  and 2~eV pronounced $d$-$d$ transitions have been identified in good agreement with simple LDA calculations, {\it i.e.}\ local density approximation without electronic correlations considered \cite{Puphal17,Pustogow17}.

\begin{figure}
\centering
\includegraphics[width=0.5\columnwidth]{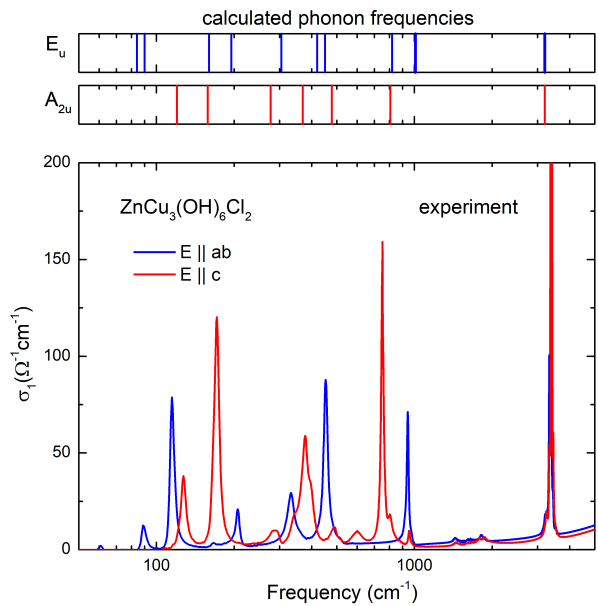}
\caption{Room-temperature optical conductivity $\sigma(\omega)$ of ZnCu$_3$(OH)$_6$Cl$_2$  measured for both polarizations, within the hexagonal $ab$-plane (blue curve) and perpendicular to it (red curve).
The observed lattice vibrations are in good agreement with the calculated phonon frequencies depicted above. Data taken from \cite{Pustogow18b}.}
\label{fig:HS-phonons}%
\end{figure}

At lower frequencies pronounced phonon features can be detected \cite{Sushkov17,Pustogow18b}, as illustrated in figure~\ref{fig:HS-phonons}. From the 54 lattice phonons the ten E$_u$ modes are infrared active for light polarized within the $ab$-plane and seven A$_{2u}$ modes for the polarization $E\parallel c$. At low temperatures, an anomalous broadening of the low-frequency phonons is observed. Sushkov {\it et al.} suggested that magnetic fluctuations in the spin-liquid state might couple to the phonons \cite{Sushkov17}. This is most pronounced in the low-frequency vibration at 115~\cm,  which involves the Cl$^{-}$ and Zn$^{2+}$ ions leading to a deformation of the kagome layer \cite{Pustogow18b}.
Raman studies reveal seven phonon modes in the corresponding spectral range \cite{Wulferding10,deVries12}. More interestingly, however, is the broad continuum attributed to scattering on magnetic excitations. When the temperature is reduced, the quasi-elastic scattering intensity decreases linearly in contrast to the exponential drop expected for a spin gap \cite{Wulferding10}.

In general, optical experiments are not sensitive to spinon excitations;
based on a U(1) gauge theory of the Hubbard model, Lee and collaborators \cite{Lee05} suggested, however, that the spin excitations in a quantum spin liquid may exhibit a Fermi surface and may also contribute to the optical conductivity \cite{Ioffe89,Ng07,Potter13} and the magneto-optical Faraday effect \cite{Colbert14}.
Hence, investigations of the electrodynamic behavior could help to
establish the spin-liquid ground state and elucidate the low-energy elementary excitations.

\begin{figure}
\centering
\includegraphics[width=0.4\columnwidth]{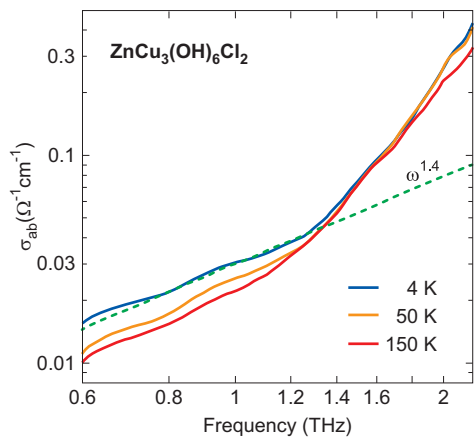}
\caption{The in-plane optical conductivity $\sigma_{ab}$ of ZnCu$_3$(OH)$_6$Cl$_2$ measured at various
temperatures as indicated. The spectra consist of a high-frequency component arising from the phonon resonance at 90~\cm, corresponding to about 2.7~THz. At low-frequencies
a power-law dependence on frequency can be identified as $\sigma(\omega)\propto \omega^{1.4}$ (dotted line). Redrawn from Ref.~\cite{Pilon13}.}
\label{fig:Pilon1}%
\end{figure}

As seen in figure~\ref{fig:HS-phonons}, besides the vibrational features
no charge excitations can be observed in the infrared spectral range, due to the large Hubbard $U$.
First THz investigations on ZnCu$_3$(OH)$_6$Cl$_2$ were performed by Gedik and collaborators \cite{Pilon13}; they report a power-law behavior of the low-frequency optical conductivity $\sigma(\omega)\propto \omega^{\beta}$ with an exponent $\beta = 1.4$ in a limited spectral range (figure \ref{fig:Pilon1}).
At 1.3~THz it crosses over to the tail of the rather strong low-frequency phonon at 90~\cm, already seen in figure~\ref{fig:HS-phonons}. The behavior is insensitive to the presence of a magnetic field ($B<7$~T), and does not change significantly when the temperature is raised to 150~K, {\it i.e.}\ well above the regime where spinon excitations are supposed to be dominant. A final conclusion can only be drawn when the observation is confirmed or supporting evidence presented. In particular, experiments at lower frequencies and temperatures are required. The confirmation of spinon contributions to the optical conductivity smoothly extending down to $\omega = 0$ could be considered as a proof of gapless excitations in kagome materials.

\section{Organic Dimerized Mott Compounds}
\begin{figure}
 \centering
 \includegraphics[width=0.6\columnwidth]{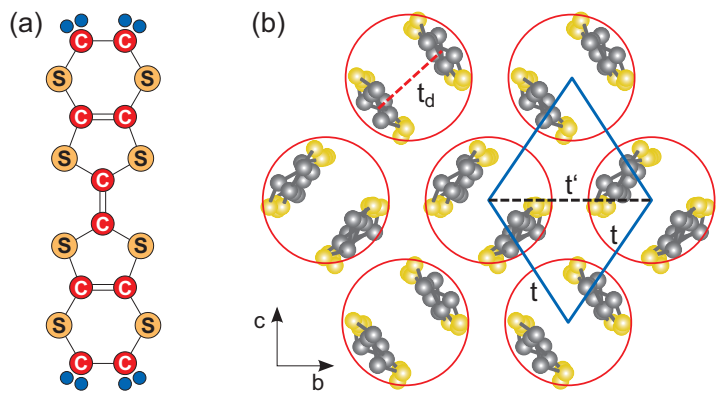}
\caption{(a)~Sketch of the
bis-(ethyl\-ene\-di\-thio)\-te\-tra\-thia\-ful\-va\-lene
molecule, abbreviated BEDT-TTF. (b)~For $\kappa$-(BEDT-TTF)$_2X$ the molecules
are arranged in dimers, which constitute an anisotropic triangular
lattice within the conduction layer. The inter-dimer transfer
integrals are labeled by $t$ and $t^{\prime}$ and can be
calculated by tight-binding studies of molecular orbitals or
ab-initio calculations \cite{Nakamura09,Kandpal09,Jeschke12}.
The intra-dimer transfer integral of \CuCN\ and \AgCN\ is
$t_d\approx 200$~meV and 264~meV, respectively.
Approximating the on-site Coulomb repulsion by $U\approx 2t_d$ \cite{McKenzie98}, one obtains at ambient
conditions $U/t=7.3$ and 10.5 with the ratio of the two inter-dimer
transfer integrals $t^{\prime}/t \approx 0.83$  and 0.90 very close to equality \cite{Nakamura09,Kandpal09,Jeschke12}.}
\label{fig:k-structure}
\end{figure}

For the molecule-based systems \CuCN\ \cite{Shimizu03,Kurosaki05},  \AgCN\  \cite{Shimizu16,Pinteric16}, and \EtMe\ \cite{Itou08,Itou10} the starting point is rather similar: here
molecular dimers with $S=\frac{1}{2}$ form a highly frustrated triangular lattice  \cite{Kandpal09,Nakamura09,Nakamura12} as sketched in figure~\ref{fig:k-structure}(b). Defining the frustration as the ratio
of transfer integrals $t^{\prime}$ and $t$, the compounds are close to perfect frustration as summarized in table~\ref{tab:1}. At ambient pressure, no indication of N{\'e}el order is observed for temperatures as low as 20~mK, despite the considerable antiferromagnetic exchange of $J\approx 220-250$~K \cite{Shimizu03,Shimizu16,Itou08}. The origin of the spin-liquid phase is unresolved since pure geometrical frustration should not be sufficient to stabilize the quantum spin-liquid state \cite{Huse88,Capriotti99,Kaneko14}.
From heat capacity measurements on \CuCN, gapless spin excitations have been concluded \cite{Yamashita08} in contrast to thermal transport data \cite{Yamashita09}.
The dispute is not completely resolved. According to Ng and Lee \cite{Ng07}, optical studies should provide important information; if spin excitations in \CuCN\ exhibit a Fermi surface, they should show up not only in the thermal conductivity but also contribute to the optical conductivity \cite{Ioffe89,Ng07,Potter13}.
Recently it was revealed, however, that disorder is intrinsic to \CuCN\ and \AgCN\ \cite{Dressel16,Pinteric16}; this aspect is crucial for understanding the electrodynamic properties of these materials in general.
\begin{table}[h]
\caption{For different molecular-based quantum spin-liquids compounds on a triangular lattice with antiferromagnetic coupling $J$ the degree of frustration $t^{\prime}/t$ is given, together with the effective correlations
defined as the ratio of inter-dimer Coulomb repulsion $U$ and bandwidth $W$; the power-law exponents are listed for the lowest temperature, according to equation~\ref{eq:powerlaw2}: $\sigma(\omega) = \sigma_0 + a \omega^{\beta_1} + b \omega^{\beta_2}$. For comparison, the respective data for Herbertsmithite are listed, which forms a kagome lattice \cite{Oshima88,Komatsu96,Shimizu16,Hiramatsu15,Itou08,Kato12a,Kato12c,Norman16,Pustogow17a}.
\label{tab:1}}
\begin{center}
{\small
\begin{tabular}{l|c c c c c}
Compound                                       &$J$ (meV) &$t^{\prime}/t$& $U/W$ & $\beta_1$ & $\beta_2$ \\
\hline
$\kappa$-(BE\-DT\--TTF)$_2$\-Cu$_2$\-(CN)$_{3}$ & 19 & 0.83 & 1.52 & 0.9 & 1.3\\
$\kappa$-(BE\-DT\--TTF)$_2$\-Ag$_2$(CN)$_{3}$   & 19 & 0.90 & 1.96 & 0.6 & 1.3 \\
$\beta^{\prime}$-EtMe$_3$\-Sb\-[Pd(dmit)$_2$]$_2$ & 22 & 0.90 & 2.35 & 1.75 & 4.2\\
\hline
ZnCu$_3$(OH)$_6$Cl$_2$                          &17 &   1    & 5-8 &1.4 & -\\
\end{tabular}}
\end{center}
\end{table}

Nakamura et al. \cite{Nakamura14} report a significant reduction in the low-energy ($\nu < 700$~\cm) Raman-scattering intensity,
when the spin-liquid compound \CuCN\ is compared to the antiferromagnetic Mott insulator \etcl.
They assign these to the two-magnon excitation processes. While in the latter compound phonon anomalies indicate the antiferromagnetic order at $T_N = 27$~K, the spin liquid compound \CuCN\ exhibit a change in the phonon lines around 90~K and a drastic drop in intensity when the compound crosses over from the thermallyßactivated semiconducting (sometimes called ``bad metallic'') to Mott-insulatoring behavior at around 50~K\cite{Rovillain14}, in agreement with comprehensive electrodynamic and transport studies \cite{Pustogow17a}.

\subsection{Infrared Properties}
\begin{figure}
 \centering
 \includegraphics[width=0.6\columnwidth]{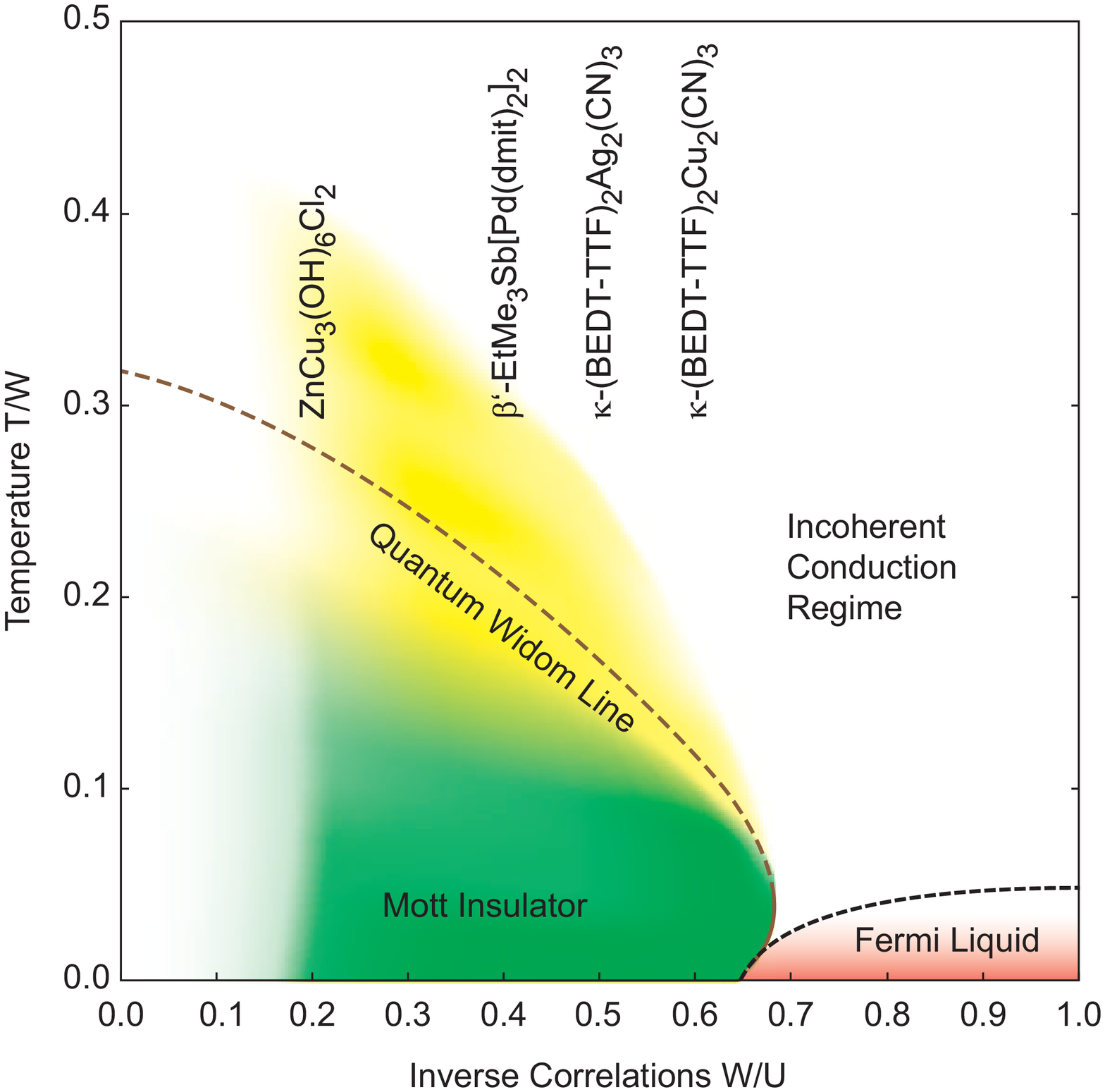}
\caption{Phase diagram of pristine Mott insulators in the absence of magnetic order.
The temperature $T$ and Coulomb repulsion $U$ are normalized by the bandwidth $W$ extracted from optical spectroscopy. The location of the Herbertsmithite ZnCu$_3$(OH)$_6$Cl$_2$,
and the dimerized Mott insulators \EtMe, \AgCN, and \CuCN\ is estimated from their optical and transport properties \cite{Pustogow17,Pustogow17a}.
Since in these quantum spin liquids, magnetic order is suppressed, the large residual entropy causes a pronounced back-bending of the quantum Widom line at low temperatures leading to metallic fluctuations in the Mott state close to the boundary. As the effective correlations decrease further, a metallic phase forms with Fermi liquid properties. }
\label{fig:phasediagram}
\end{figure}

\begin{figure}
\centering
\includegraphics[width=0.8\columnwidth]{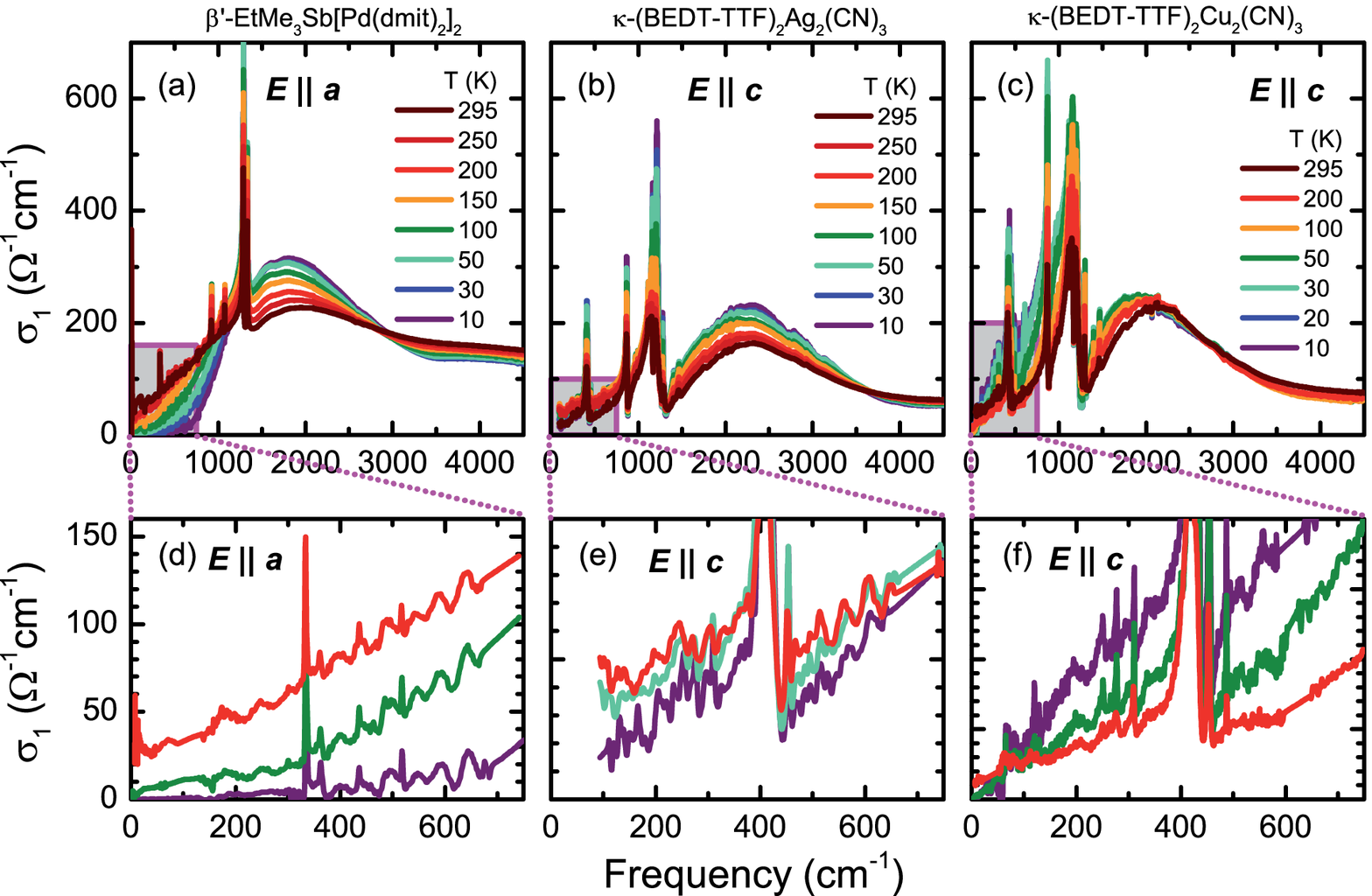}
\caption{Temperature evolution of the optical conductivity of \EtMe, \AgCN\ and \CuCN\ within the conducting planes as indicated. (a)-(b) The pronounced band in the mid-infrared range is associated with the excitation between the Hubbard bands. The peak position is a measure of the Coulomb repulsion $U$. The lower panels (d)-(f) enlarge the low frequency part in order to demonstrate the qualitative differences in the temperature dependence strongly depending on the position in the phase diagram. While the optical conductivity $\sigma(\omega)$
decreases upon cooling for the more correlated \EtMe\ and \AgCN, it shows a pronounced non-thermal enhancement for \CuCN\ situated most closely to the Mott critical point. The data are taken from \cite{Pustogow17a} and discussed there in full detail.
\label{fig:organics_cond}}
\end{figure}

In contrast to the completely insulating Herbertsmithite, where the on-site Coulomb repulsion $U$ is  large compared to the hopping integral $t$, the electron-electron repulsion for the dimerized charge-transfer salt is significantly lower: $U\approx 0.5$~eV. As illustrated in figure~\ref{fig:phasediagram}, \CuCN\ is very close to the Mott insulator-to-metal transition ($U \approx 1.5\, W$) \cite{Pustogow17a}; in fact under hydrostatic pressure of only 4~kbar it becomes superconducting at $T_c=3.6$~K \cite{Geiser91,Shimizu03}. For \AgCN\ approximately 10~kbar are needed to enter the superconducting state \cite{Shimizu16}.
In figure~\ref{fig:organics_cond} we display the infrared conductivity
for three organic spin-liquid candidates measured at various temperatures using light polarized
within the highly conducting plane. The overall shape of  $\sigma(\omega)$ is rather similar and dominated by a broad band that peaks around 1700 - 2300~\cm, which arises from the electronic transition between the lower and upper Hubbard bands and is therefore a measure of the effective Coulomb repulsion $U$ \cite{Faltermeier07,Merino08,Dumm09,Dressel09,Ferber14}.
The values for $U/W$ extracted from figure~\ref{fig:organics_cond} are listed in table~\ref{tab:1} \cite{Pustogow17a}.

Besides these electronic excitations, sharp vibrational features are observed below 1500~\cm. In particular for the $\kappa$-phase BEDT-TTF salts the A$_g$ intramolecular vibrations -- supposed by be infrared silent -- become infrared active by electron-molecular vibrational (emv) coupling to the charge-transfer excitations. They are shifted down in frequency with respect to the corresponding Raman
modes \cite{Girlando00,Dressel04,Yakushi12,Girlando12,Girlando14} and are strongly distorted.
Numerical investigations of the electronic ground state and the molecular and lattice vibrations
reveal the importance of the anion network for the vibrational features extending down to the THz range of frequency \cite{Dressel16,Pinteric16,Pinteric18}.

In order to elucidate the low-energy excitations within the Mott gap, the lower frames in figure~\ref{fig:organics_cond} enlarge the optical conductivity in the far-infrared spectral range. The three compounds exhibit a rather different temperature evolution. For \EtMe, the compound with strongest effective correlations $U/W$, a clear Mott gap opens around 125~K and continuously grows up to 650~\cm\ when the temperature is lowered to 5~K. Also \AgCN\ behaves like a typical insulator, where the in-gap states are depleted upon cooling \cite{Nakamura17}. The opposite behavior is found in the case of \CuCN\ [figure~\ref{fig:organics_cond}(f)], which exhibits a temperature dependence characteristic of metals: the low-frequency optical conductivity increases upon lowering the temperature \cite{Kezsmarki06,Sedlmeier12,Elsasser12}. The far-infrared spectral weight of the weakly correlated spin-liquid compound \CuCN\ increases significantly for $T<100$~K and
no clear-cut Mott gap is observed in the optical spectra.
This is unexpected considering that no Drude peak is present -- the hallmark of coherent transport -- and
that at zero frequency all three compounds -- including \CuCN\ -- are electrical insulators as determined from dc transport \cite{Pustogow17a}.

\subsection{$\kappa$-(BE\-DT\--TTF)$_2$\-Cu$_2$\-(CN)$_{3}$}
\begin{figure}[h]
\centering
\includegraphics[width=0.5\columnwidth]{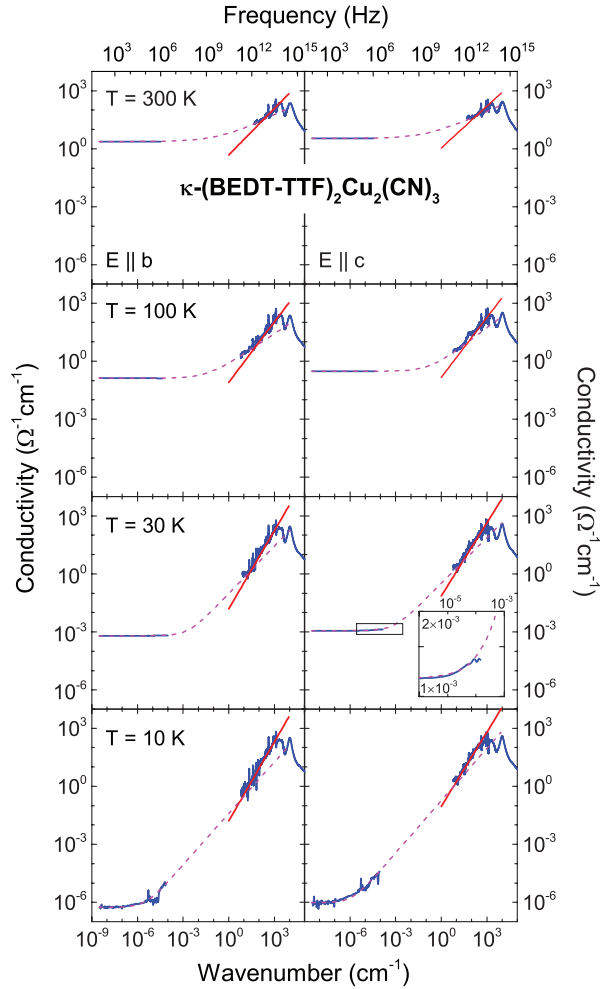}
\caption{Overall optical conductivity of \CuCN\ for the two polarizations $E\parallel b$ (upper row) and $E\parallel c$ (lower row) for several temperatures as indicated. The dashed magenta lines are fits
by Eq.~(\ref{eq:power})
to connect the two ranges, extending to a power-law behavior in the THz range.
The solid red lines indicate the high-frequency power law. The inset is an enlargement of the frame in order to demonstrate the conductivity rise in the MHz range. The data are taken from \cite{Dressel16,Pinteric14,Pustogow14}.
\label{fig:Conductivity}}
\end{figure}

\begin{figure}
\centering
\includegraphics[width=0.5\columnwidth]{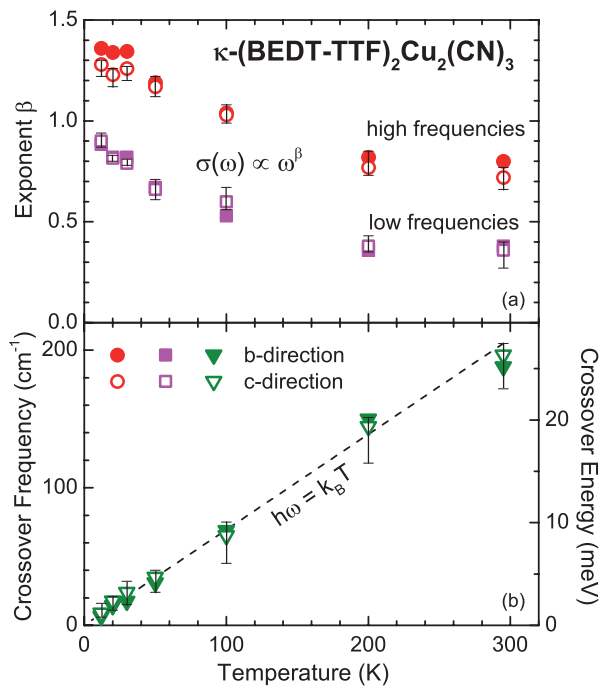}
\caption{(a)~Temperature dependence of the power-law exponents $\sigma_1(\omega)\propto \omega^\beta$ of \CuCN\
for $E\parallel b$ (solid symbols) and $E\parallel c$ (open symbols).
(b)~The crossover frequency $\omega_c$ between both regimes shifts with temperature
corresponding to $\hbar\omega_c\approx k_BT$. The error bars (shown only for the $c$-polarization) indicate the uncertainty
in the power-law fit and the determination of the crossover frequency, respectively.
\label{fig:Exponent}
}
\end{figure}

In order to gain more insight into the low-energy dynamics, in figure~\ref{fig:Conductivity} the conductivity $\sigma(\omega)$ of \CuCN\ --~as the best studied example of the organic quantum spin-liquid candidates~-- is displayed in a wide range of frequencies for both in-plane polarizations and different temperatures between $T=300$ and 10~K.
In the kHz and MHz range, hopping conduction is identified as the dominant transport mechanism, accompanied by a broad dielectric relaxation at lower frequencies that bears typical fingerprints of relaxor ferroelectricity \cite{Abdel-Jawad10,Pinteric14,Pinteric14b,Tomic15,Dressel16}. For low temperatures, $T<50$~K,
we find an appreciable increase in $\sigma(\omega)$ of the high-frequency dielectric data, which nicely matches the slope observed in the GHz and THz range.
The dashed magenta lines in figure~\ref{fig:Conductivity} simply interpolate the break in our data by
\begin{equation}
\sigma(\omega)=\sigma_0 + a \omega^\beta \quad ,
\label{eq:power}
\end{equation}
with a temperature-dependent constant $\sigma_0$ and prefactor $a$. The exponent $\beta$
of the power law is approximately $0.4$ and increases to almost 1 when the temperature is reduced below $T=100$~K, as summarized in figure~\ref{fig:Exponent}(a). It should be noted that the rise
in $\sigma(\omega)$ and the corresponding power-law behavior is already observed above  300~kHz in the low-temperature dielectric data (figure~\ref{fig:Conductivity}). Covering such an extremely broad spectral range leads to a high confidence in the power-law exponents \cite{Pustogow14}.

Interestingly, in figure~\ref{fig:organics_cond} we do not observe a gradual filling or closing of the gap in \CuCN\  but an increase in  the exponent $\beta$ of the power-law $\sigma(\omega)\propto \omega^{\beta}$ as the temperature is reduced.
Our robust observation \cite{Kezsmarki06,Elsasser12} is an unambiguous signature that these  low-energy excitations are of quantum and not of thermal origin.
The opposite behavior is found the case of the antiferromagnetic Mott insulator \etcl, where the Mott gap gradually fills as chemical pressure is applied by Br substitution, leading to a reduction of the effective Coulomb repulsion $U/W$ \cite{Faltermeier07,Dumm09}. For comparison,
pressure-dependent optical studies  \cite{Beyer16} could show, that
the charge-order gap in $\alpha$-(BEDT-TTF)$_2$I$_3$ exhibits a linear suppression of the gap as the hydrostatic pressure increases.

\begin{figure}
\centering
\includegraphics[width=0.9\textwidth]{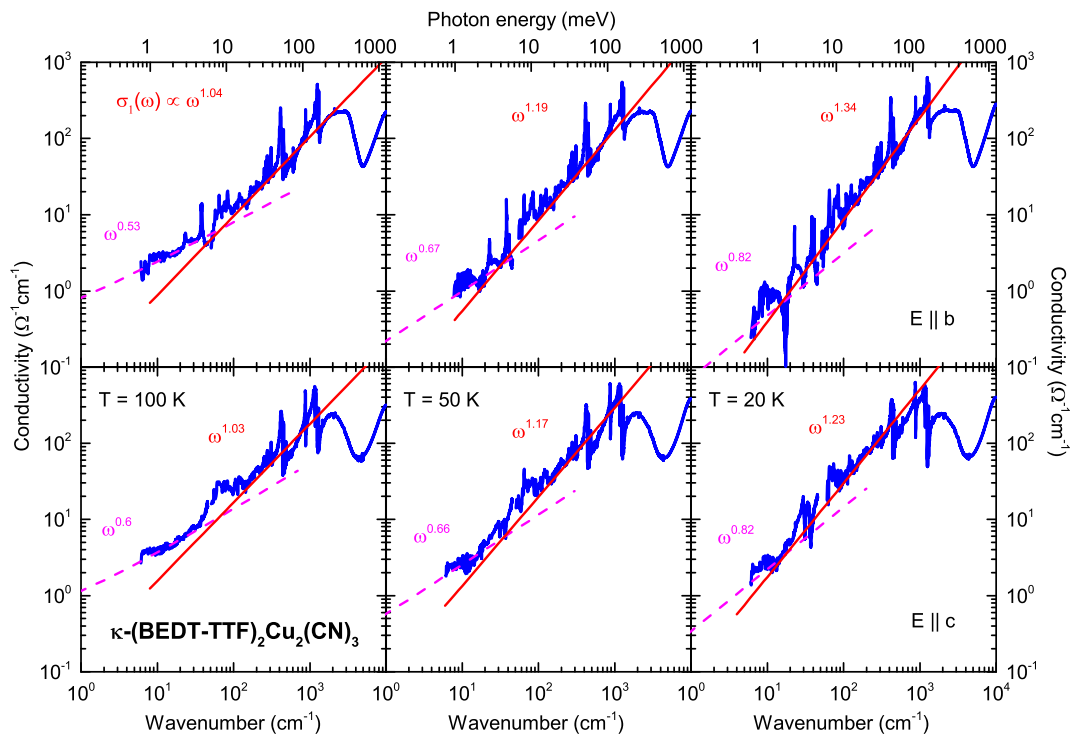}
\caption{Optical conductivity of \CuCN\ measured at $T=100$, 50, and 20~K for the two crystallographic directions, upper and lower panels as indicated. The dashed magenta lines correspond to the power-law behavior extending to low-frequencies, the solid red lines indicate the high-frequency power-laws:
$\sigma = \sigma_0 +a\omega^{\beta 1} + b \omega^{\beta 2}$. The data are taken from \cite{Dressel16,Pustogow14}.
\label{fig:PowerLaw}}
\end{figure}

Similarly intriguing is the change of the power-law behavior in the THz frequency range. In figure~\ref{fig:PowerLaw} the data of the electrodynamic response of \CuCN\ are re-plotted on a magnified scale for some selected temperatures. At $T=100$~K, for example, the slope clearly changes around $\nu_c=\omega_c/2\pi c \approx 100$~\cm, where the exponent increases from $\beta=0.53$ to 1.04 for $E\parallel b$ (upper panel) and from 0.60 to 1.03 for $E\parallel c$ (lower panel). This crossover can be identified for all temperatures and both polarizations \cite{Pustogow14,remark1}. We summarize our finding on \CuCN\ in figure~\ref{fig:Exponent}.
It should be noted that the high-frequency slope is not affected by a phonon tail that obscures the data in the Herbertsmithite \cite{Pilon13}, as demonstrated in figure~\ref{fig:Pilon1}.

At first glance our findings seem to agree well with the suggestion of
Ng and collaborators \cite{Ng07,Zhou13} that un-gapped spinons at the Fermi surface
contribute to the optical behavior of a quantum spin liquid.
They predict a strongly enhanced conductivity
within the Mott gap compared to the two spin wave absorption in a N{\'e}el-ordered insulator.
From this point of view, the significant increase in conductivity at low temperatures
could support the conclusion of gapless spin excitations.
In addition, the calculations yield a power-law absorption at low frequencies, {\it i.e}.\
for energies smaller than the exchange coupling $J$.
For very low energies ($\hbar\omega < k_BT$), the
optical conductivity $\sigma(\omega)\propto\omega^2$, and for
$\hbar\omega
> k_BT$ the power law should increase to $\sigma(\omega) \propto
\omega^{3.33}$ if impurity scattering was negligible \cite{Ng07}.
While we do observe a kind of crossover frequency close to thermal energy, $\beta_1$ and $\beta_2$ are consistently lower
by more than a factor of 2.
We want to recall that also for the Herbertsmithite ZnCu$_3$(OH)$_6$Cl$_2$ the
power-law exponent extracted in the low-frequency range was only $\beta\approx 1.4$ \cite{Pilon13}, as displayed in figure~\ref{fig:Pilon1}.

Here we might speculate whether this is
influenced by inherent inhomogeneities in the system \cite{Mendels07,Helton07,Pinteric16,Dressel16}: when localization effects dominate, the electronic system behaves like a disordered metal \cite{Ng07} and exhibits the corresponding frequency dependence \cite{MottDavis79}.

\subsection{Metallic Quantum Fluctuations}
\begin{figure}[b]
\centering
\includegraphics[width=0.8\columnwidth]{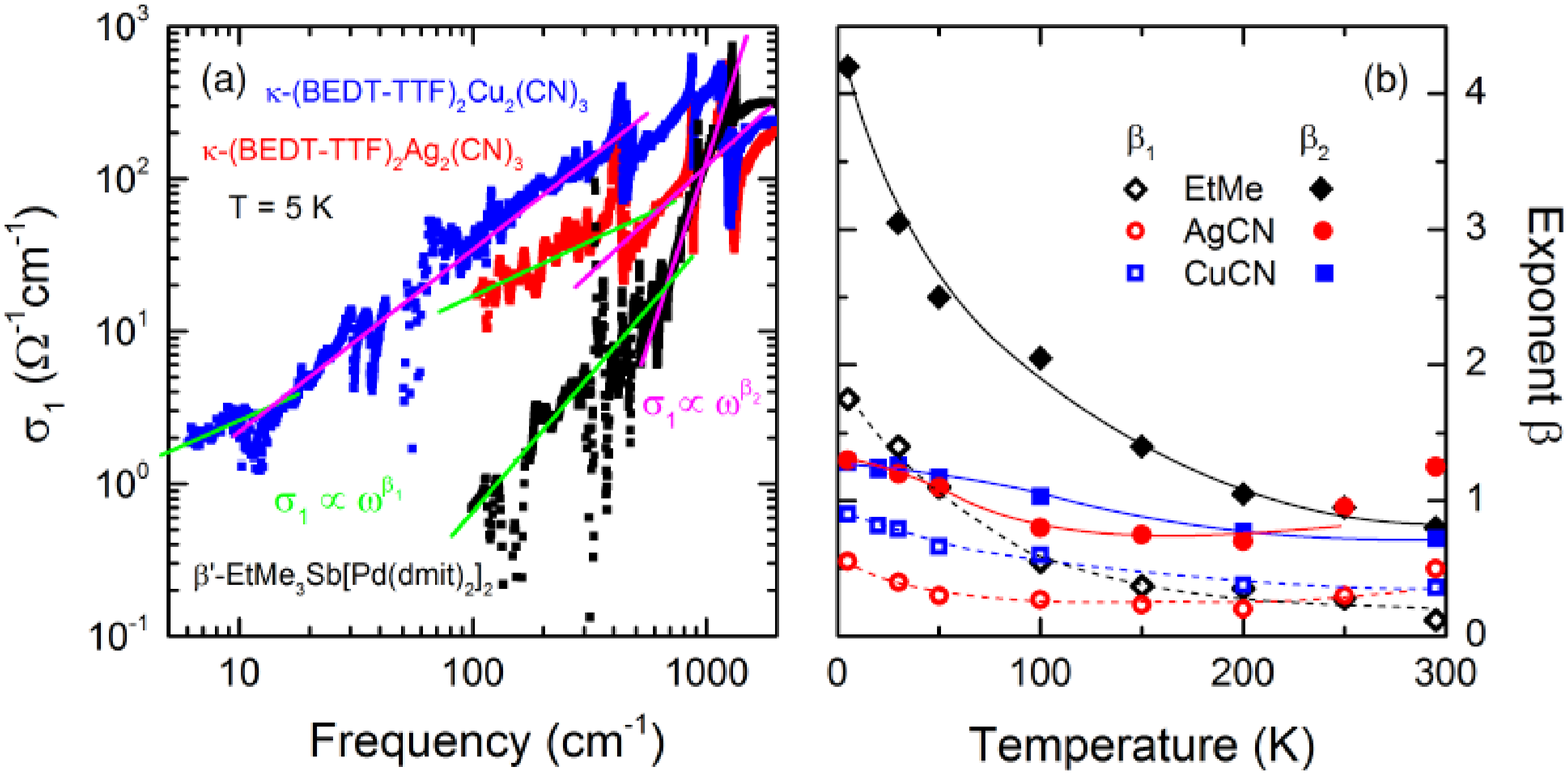}
\caption{(a)~At the smallest accessible temperature, the low-frequency optical conductivity
of the organic spin-liquid Mott insulators
can be described by power laws $\sigma(\omega)= \sigma_0 + a \omega^{\beta_1} + b \omega^{\beta_2}$. For all three compounds, \EtMe\ ($E\parallel a$, black symbols),  \AgCN\ ($E\parallel c$, red symbols) and  \CuCN\ ($E\parallel c$, blue symbols), two distinct ranges can be identified:
a smaller exponent at low frequencies with a crossover to about twice the value at high frequencies. For the systems \AgCN\ and \CuCN\ the low temperature exponent does not exceed unity at low frequencies, while it is enhanced to 1.8 in the case of \EtMe. (b) The power-law exponents $\beta$ in general increase as the temperature is reduced and the Mott gap opens upon cooling. This is more pronounced for the strongly correlated \EtMe. Data taken from \cite{Pinteric14,Pinteric16,Pustogow17a}
\label{fig:allpowerlaws}
}
\end{figure}

The corresponding analysis can be conducted for all three quantum-spin liquid candidates in order to extract the general tendency and unveil the dependence on the material parameters.
In figure~\ref{fig:allpowerlaws}(a) the frequency-dependent conductivity of \EtMe, \AgCN\ und \CuCN\
is summarized for the lowest temperature ($T = 10$~K) measured perpendicular the respective $b$-direction. In the double-logarithmic presentation the power-law behavior
\begin{equation}
\sigma(\omega) = \sigma_0 + a \omega^{\beta_1} + b \omega^{\beta_2}
\label{eq:powerlaw2}
\end{equation}
becomes obvious with two distinct exponents $\beta_1$ and $\beta_2$;
their temperature evolution is plotted in figure~\ref{fig:allpowerlaws}(b).
The behavior is qualitatively similar for the polarization $E\parallel b$,
indicating no significant anisotropy for the in-plane electronic properties,
as already seen from figure~\ref{fig:Exponent}.
Upon cooling, the in-gap states freeze out, the dc conductivity drops,
and the slope enhances correspondingly.
For \CuCN\ and even \AgCN\ the temperature dependence is much weaker compared to the large changes observed in \EtMe, where it resembles the formation of the Mott gap, similar to the results shown in figure~\ref{fig:organics_cond}.
While the high-frequency exponent reaches a comparable value for \CuCN\ and \AgCN\
at the lowest temperature ($\beta_2 \approx 1.4$), it approaches $\beta_2  \approx 4.5$ for \EtMe.

In a recent theoretical study Dobrosavljevi{\'c} and collaborators \cite{Lee16} concluded that spinons are not well-defined close to the Mott transition. As soon as the
electrons become delocalized, the spins have to follow the charge movement, which
destroys the coherence of the postulated spinon Fermi surface.
Thus spinons do not play a significant role within the high-temperature quantum critical regime above the Mott point, in contrast to previous suggestions \cite{Lee05,Senthil08,Podolsky09}.
In other words, fingerprints of spinon excitations can only be expected in the
optical properties of those spin-liquid compounds, which are located deep inside the Mott state,
making sure that charge response is completely absent.
For the examples of the three organic crystals discussed here,
\EtMe\ is the prime candidate to find evidence for spinon contributions to the conductivity,
according to table~\ref{tab:1} and figure~\ref{fig:phasediagram}. Even then we have to look at rather small frequencies and temperatures well below the antiferromagnetic exchange coupling $J\approx 250$~K \cite{Itou08}.

\section{Spinon Contribution}
\begin{figure}
\centering
\includegraphics[width=0.7\columnwidth]{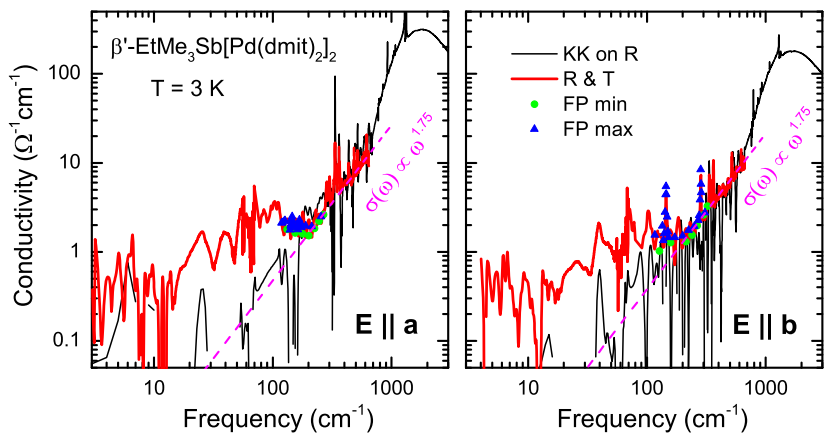}
\caption{The low-frequency conductivity of \EtMe\ determined using
different methods at $T=3$~K for two polarizations of the electric field.
Apart from the absolute values, there is no
appreciable anisotropy between the two in-plane crystal axes $a$ and $b$.
The data obtained from the Kramers-Kronig analysis of the reflectivity  (thin black line) become rather noisy
below 100~\cm\ and contains an increasing uncertainty due to the required extrapolation.
The results directly obtained from transmission and reflection measurements (thick red line) are more reliable. In addition, fits of the Fabry-Perot interference maxima and minima enable us to directly calculate the optical conductivity (green circles and blue triangles).
From 200 to 500~\cm\ a power law can be fitted to the data, which crosses over to a larger slope reminiscent of the Mott-Hubbard band.
For $\nu < 200$~\cm, however, $\sigma_1$ exceeds the extrapolated power-law
descent towards lower frequencies indicative of an additional absorption mechanism.
\label{fig:Cond1}
}
\end{figure}

By performing THz transmission measurements on \EtMe\
single crystals of various thicknesses at temperatures as low as 3~K,
it was possible to cover the range down to 3~\cm\ with
the required accuracy and directly evaluate the optical conductivity without
any extrapolation \cite{Pustogow18a}.
As seen from figure~\ref{fig:Cond1} for both polarizations
within the highly plane, an additional absorption process
can be identified that adds to the electronic background \cite{Pustogow18a}.
According to figure~\ref{fig:allpowerlaws}
the underlying conductivity solely due to charge excitations
follows a power law $\sigma_1(\omega)\propto \omega^{1.75}$ in the far-infrared range,
crossing over to a steeper slope where the tail of the Mott-Hubbard band dominates.

Below approximately 200~\cm, however,  the optical conductivity exceeds the
power-law extrapolation considerably,
giving evidence for an additional contribution to the electrodynamic response.
At much lower frequencies, $\sigma_1(\omega)$ gradually levels off
toward the dielectric data \cite{Lazic18},
comparable to what is plotted in figure~\ref{fig:Conductivity} for \CuCN.
The transition regime is consistent with a $\sigma_1(\omega)\propto \omega^2$
behavior as suggested by Ng and Lee \cite{Ng07},
that catches up the interpolated value of the correlated electrons rather quickly.
Although for $T\rightarrow 0$ this range is enlarged to lower energies,
as hopping conduction freezes out, the $\omega^{1.75}$ decay
is approached asymptotically in the GHz range.
This gives the lower bound of the spinon-dominated optical conductivity.
Thus, the realm of coherent spinons is limited to the range from MHz frequencies to 200~\cm.
When the electronic conductivity exceeds the spinon contribution, the related
Fermi surface is eventually damped away, as discussed in Ref.~\cite{Lee16}.

\begin{figure}
\centering
\includegraphics[width=0.5\columnwidth]{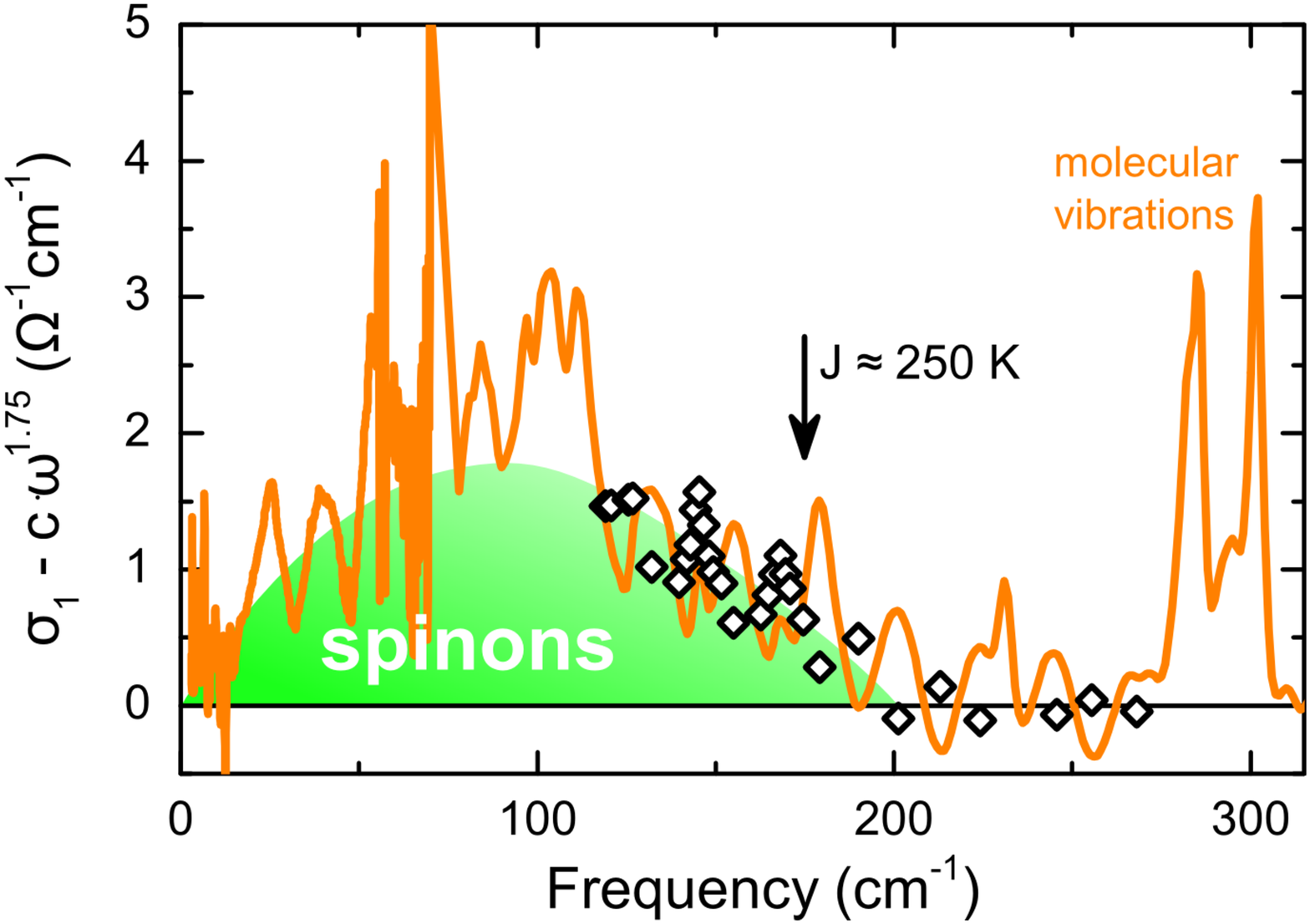}
\caption{After subtracting the power-law background related with the Mott-Hubbard
band, we obtain a broad low-energy feature below approximately 200~\cm,
which is close to the antiferromagnetic exchange energy of \EtMe.
Hence, we associate it with the coherent spinon Fermi surface predicted
previously \cite{Lee05,Ng07}. The dome-like shape of the spinon
contribution arises due to the $\omega^2$ decay towards $\omega \rightarrow 0$.
This feature becomes visible when
the background conductivity is sufficiently suppressed due to opening of the Mott gap. The
strong, narrow modes at higher frequencies correspond to vibrational features.
\label{fig:Cond3}
}
\end{figure}

Now we can subtract the smooth electronic background determined in figure~\ref{fig:Cond1}
and focus on the excess conductivity, that is  seen in the linear plot of figure~\ref{fig:Cond3} as a broad feature of the optical conductivity below 200~\cm.
Apart from a few vibrational features, it is rather isotropic and confined to a frequency range determined by the antiferromagnetic exchange $J \approx 250~{\rm K} = 175$~\cm.
This dome-shaped in-gap absorption is naturally attributed to spinons,
which occur when $J$ is the dominant energy scale and the electronic conductivity
is sufficiently suppressed at low temperatures.
In the static limit, $\sigma_1(\omega)$ decays faster
than the $\omega^{1.75}$ power-law background of the optical data.
Thus spinons affect neither the optical range nor the dc response where the physics of correlated electrons prevails; nevertheless they can be observed at finite temperatures
in a limited frequency range  well below the Mott gap.

Going back to the overview on several quantum spin liquids plotted in figure~\ref{fig:phasediagram}, we can now understand why
for \CuCN\ no indications of spinons could be seen in the optical conductivity (figure~\ref{fig:PowerLaw}) \cite{Kezsmarki06,Elsasser12}. Due to the weaker correlations $U/W$ the compound is located much closer to the insulator-to-metal phase boundary \cite{Pustogow17a} and consequently the tail of the Mott-Hubbard excitations decay much slower towards $\omega\rightarrow 0$. \CuCN\ exhibits a power-law conductivity with a weaker slope and larger absolute value compared to \EtMe. Hence, the electronic contribution to the electrodynamic response of \CuCN\ dominates well into the GHz range of frequency.

\section{Summary}
In all cases that came under scrutiny here, the power laws observed in all spin-liquid compounds are not in accordance with the theoretical values predicted for spinon contributions \cite{Ng07}, neither at low, nor at high frequencies. In the low-frequency limit, the exponent $\beta_1$ generally resembles Jonscher's universal power law of the dielectric response \cite{Jonscher77} as it is widely observed in disordered solids. At high frequencies and temperatures spinons should not play a role because the excitation energy exceeds the antiferromagnetic exchange coupling $J$. The frequency and  temperature dependence observed in the optical conductivity of these quantum spin liquid compounds is governed by charge excitations rather than magnetic contributions.
For systems with a large Coulomb gap, such as \EtMe\ or Herbertsmithite,
the spinon contribution becomes detectable in the GHz  and THz ranges at very low temperatures when the in-gap absorption is significantly reduced. Here we can identify a contribution to the $\omega\rightarrow 0$ optical conductivity that can be assigned to gapless spinon excitations.

\ack
Over the years, we experienced fruitful collaborations with several groups in the various fields.
We acknowledge the fruitful collaboration with and contributions from V. Dobrosavljevi{\'c}, S. Fratini, B. Gorshunov, R. H{\"u}bner, T. Ivek, R. Kato, C. Krellner, Y. Li, A. L{\"o}hle, I. Mazin, M. Pinteri{\'c}, P. Puphal, R. R{\"o}sslhuber, G. Saito, J.A. Schlueter, R. Valent{\'i}, S. Tomi{\'c}, Y. Yoshida and E. Zhukova. The work was supported by the Deutsche Forschungsgemeinschaft (DFG) via DR228/39-1 and DR228/52-1.

\section*{References}

\begin{thebibliography}{100}
\bibitem{remark1}
The extracted values are in accord with the preliminary analysis of the
  independent experiments on a different \CuCN\ single crystal presented in
  Ref.~\cite{Elsasser12}.
\end{thebibliography}
\providecommand{\newblock}{}

\end{document}